\documentclass{ws-procs9x6}

\newcommand{\mbf}[1]{\mbox{\boldmath $#1$}}

\begin{document}

\title{Color superconductivity from magnetic interaction induced
by flow equations}

\author{E. GUBANKOVA\footnote{
Present address: Center for Theoretical Physics,
Massachusetts Institute of Technology, Cambridge, MA 02139.
E-mail: elena1@mit.edu}}

\address{Institute of Theoretical and Experimental Physics,\\
B. Cheremushkinskaya 25, RU-117 218\\
Moscow, Russia}

\maketitle

\abstracts{Using flow equations, we derive an effective quark-quark 
interaction and obtain the coupled set of gap equations
for the condensates of the CFL phase of massless $N_f=3$
dense QCD. We find two different sources of the infrared
cutoff in magnetic interaction. When the penetration depth
of the magnetic field inside the superconductor is less than
the coherence length of quark-quark bound state, 
our results for the gap agree with those of Son \cite{Son}. 
In the other case, we obtain parametric enhancement of the gap
on the coupling constant.}

\section{Introduction}

Quark matter at sufficiently high density is a color superconductor.
It has been known for some time \cite{old}, however the value of 
the gap was too small to be of any practical interest. Recent
revision of the subject brought much larger gap \cite{revised} than 
previously thought. The consistent numerical value of the gap has been obtained
using phenomenological four-fermion interactions
\cite{AlfordRajagopalWilczek} and from QCD one-gluon exchange \cite{Son},
though with different parametric dependence on the coupling.
Many microscopic calculations of the gap in $N_f=2$ \cite{N_f=2,N_f=2more}
and in $N_f=3$ \cite{N_f=3} have been done after that.

Microscopic studies rely on the fact that due to asymptotic freedom
QCD becomes a weakly interacting theory at high densities and
perturbation theory can be used. However, high density quark matter
cannot be described as a simple Fermi gas. As known from the BSC
theory of superconductivity, the Fermi surface is unstable to
an arbitrarily weak attractive interaction. In QCD attractive 
interaction is mediated by one-gluon exchange between quarks
in color antitriplet state.

In the language of renormalization group, BCS instability
with respect to Cooper pairing is associated with breaking down
the conventional perturbation theory in the vicinity 
of the Fermi surface. This signals formation of a gap which
regulates the infrared divergent behavior. It is in full analogy
with electron superconductivity in metals. The difference between
the two is that the attractive interaction via gluon exchange
is already present in QCD rather than being induced by the 
solid state lattice as in metal. Therefore, properties of a gluon
propagating in high density quark matter are essential in calculating
the color superconducting gap. 
It has been suggested by Son (the first reference in \cite{Son}), 
that the gap is dominated by magnetic long range exchanges, 
and that the infrared behavior
is regulated by dynamic screening in quark plasma. We show
that there are two characteristic scales in superconducting medium:
the penetration depth of magnetic field inside the superconductor,
$\delta$, and the coherence length of the bound quarks (or size of the
$qq$ Cooper pair), $\xi$. The prevailing energy $1/\delta$ or $1/\xi$
regulates the long range behavior of magnetic field 
in color superconductor.

The purpose of the present work is to obtain an effective 
microscopic low energy theory of quasiparticles and holes 
in the vicinity of the Fermi surface. We shall argue that depending on 
the state of superconductor, there are two different sources of the
infrared cutoff in magnetic interaction. At small temperature,
the superconductor is of a Pipperd type, $\delta\ll\xi$, and magnetic gluon
is damped by the quark medium according to Landau damping mechanism.
As we increase the temperature, the quark plasma 'dilutes'
and the penetration depth of magnetic field grows. It turns out
that usually the superconducting condensate melts slower than
the penetration depth grows \cite{Abrikosov}, corresponding to London type 
of superconductor, $\xi\ll\delta$. In this case magnetic gluon scatters
over diquark Cooper pairs and damped by the superconducting medium.
Magnetic interaction is regulated by the inverse coherence length,
or roughly by the superconducting gap. In these two limiting cases
we obtain for the gap the same parametric dependence on coupling constant,
though different numerical factors.

We apply flow equations to the Coulomb gauge QCD Hamiltonian with
$N_f=3$ at nonzero quark density. Flow equation method \cite{Wegner}, 
which is a synthesis of the perturbative renormalization group 
and the many-body technique, has been successfully applied to 
the problem of electron superconductivity in metals \cite{LenzWegner}. 
The main idea of the method is to eliminate interactions
which mix states with different particle content (or number of particles)
and obtain an effective theory which conserves particle number
in each (Fock) sector. This procedure is performed in sequence for
matrix elements with large energy differences down to more degenerate
states. In high density QCD, all nonabelian contributions are suppressed,
therefore, the quark-gluon coupling is the only term which mediates
hoping between states of high- (away from Fermi surface) and low-
(near Fermi surface) energies. Eliminating quark-gluon coupling
by flow equations we virtually move towards the Fermi surface,
in analogy to perturbative renormalization group scaling.
Decoupling different energy scales corresponds to integrating out
fast moving modes - quarks in the Dirac sea, and obtaining an effective
theory for slow modes - quarks close to Fermi surface. 
As mentioned above, in this scaling down we encounter two 
characteristic scales: penetration depth of magnetic field and
the coherence length of $qq$ bound state. Depending which scale
is bigger, we obtain different mechanisms of the infrared cutoff
in magnetic interaction and corresponding gaps.

\section{Effective microscopic Hamiltonian for color superconductivity}
\label{effectiveHamiltonian}

We start with the Coulomb gauge QCD Hamiltonian,$\nabla\cdot A =0$, with
$N_f=3$ at nonzero quark densities given by
\begin{eqnarray}
H &=& H_{0}+H_{I}=H_{0}+H_{inst}+H_{dyn}
\,,\label{eq:2.1}\end{eqnarray}
where $H_{0}$ is the free Hamiltonian, $H_{inst}$ is the instantaneous interaction
describing static properties, and $H_{dyn}$ is the dynamical interaction 
involving the gluon propagation. The free part is the kinetic energy
of quarks and gluons
\begin{eqnarray}
 H_{0} &=& \int d \mbf{x}\bar{\psi}(\mbf{x}) 
\left( -i \mbf{\gamma}\cdot\mbf{\nabla} -\mu\gamma_0
+ m \right)\psi(\mbf{x})
\nonumber\\ 
&+& {\rm Tr}\int d \mbf{x} \left( \mbf{\Pi}^2(\mbf{x})
+ \mbf{B}^2_{A}(\mbf{x}) \right)
\,,\label{eq:2.2}\end{eqnarray}
where the non-abelian magnetic field is
$\mbf{B}=B_i=\nabla_j A_k-\nabla_k A_j+g[A_j,A_k]$, and 
its abelian part is represented by $\mbf{B}_{A}$.
The degrees of freedom are the transverse perturbative gluon field 
$\mbf{A}=\mbf{A}^{a}T^a$ ($A\equiv A^{PT}$), 
its conjugate momentum $\mbf{\Pi}$, and the quark
field in the Coulomb gauge. 
The instantaneous interaction is given by
\begin{eqnarray}
 H_{inst} &=& -\frac{1}{2}\int d\mbf{x}d\mbf{y}
\bar{\psi}(\mbf{x})\gamma_0 T^a\psi(\mbf{x})
V_{inst}(|\mbf{x}-\mbf{y}|)
\bar{\psi}(\mbf{y})\gamma_0 T^a\psi(\mbf{y})
\,,\label{eq:2.3}\end{eqnarray}   
where the leading order kernel is a Coulomb potential
defined, together with its Fourier transform, by 
\begin{eqnarray}
V_{inst}(r) &=& - \frac{\alpha_s}{r}~,~
V_{inst}(q) = - \frac{g^2}{\mbf{q}^2}
\,,\label{eq:2.4}\end{eqnarray}
with $\alpha_s=g^2/4\pi$.
The dynamical interaction includes the minimal quark-gluon coupling,
$V_{qg}$, and the non-abelian three- and four-gluon interactions,
$V_{gg}$, i.e. $H_{dyn} = V_{qg} + V_{gg}$, where
\begin{eqnarray}
 V_{qg} &=& -g \int d \mbf{x}\bar{\psi}(\mbf{x})
\mbf{\gamma}\cdot\mbf{A}(\mbf{x}) \psi(\mbf{x})
\nonumber\\
 V_{gg} &=& {\rm Tr}\int d \mbf{x}
\left(\mbf{B}^2(\mbf{x})-\mbf{B}^2_{A}(\mbf{x})
\right)
\,.\label{eq:2.5}\end{eqnarray}
As mentioned in the introduction, at high quark densities 
diagrams including the non-abelian interactions are suppressed, 
henceforth $V_{gg}$ is not considered. 

Following the work of Alford, Rajagopal and Wilczek,
\cite{AlfordRajagopalWilczek}, we assume,
due to quark interactions, the diquark condensation
in scalar $<\psi^{T}C\gamma_5\psi>$ and pseudoscalar
$<\psi^{T}C\psi>$ channels, or equivalently
$<bb>\neq 0$ and $<b^{\dagger}b^{\dagger}>\neq 0$.
In other words, perturbative vacuum $|0\rangle$
is not a ground state for the system, instead it is 
the BCS vacuum $|\Omega\rangle$ containing condensates of diquarks.
The Fock space is constructed from this vacuum using
quasiparicle operators $b^{\dagger}$ and $d^{\dagger}$ 
\begin{eqnarray}
&& \psi(\mbf{x}) = \sum_{s}\int\frac{d\mbf{k}}{(2\pi)^3}
    [u(\mbf{k},s)b(\mbf{k},s)
    +v(-\mbf{k},s)d^{\dagger}(-\mbf{k},s)] {\rm e}^{i\mbf{k}\mbf{x}} 
    \nonumber\\
&& \mbf{A}(\mbf{x}) = \sum_{a}\int\frac{d\mbf{k}}{(2\pi)^3}
\frac{1}{\sqrt{2\omega(\mbf{k})}}
    [a(\mbf{k},a)+a^{\dagger}(-\mbf{k},a)] {\rm e}^{i\mbf{k}\mbf{x}} 
    \nonumber\\
&& \mbf{\Pi}(\mbf{x}) = -i\sum_{a}\int\frac{d\mbf{k}}{(2\pi)^3}
\sqrt{ \frac{\omega(\mbf{k})}{2} }
    [a(\mbf{k},a)-a^{\dagger}(-\mbf{k},a)]{\rm e}^{i\mbf{k}\mbf{x}} 
\,,\label{eq:2.6}\end{eqnarray}
where $b|\Omega\rangle=d|\Omega\rangle =0$, and the gluon part
has trivial vacuum $a|0\rangle =0$ with $\omega(\mbf{k})=|\mbf{k}|$.
All descrete numbers 
(helicity, color, and flavor for the quarks and 
color for the gluons) are collectively denoted as $s$ and $a$, respectively.
The gluon operators $a=a^{IA}(\mbf{k})=
\sum_{\lambda=1,2}\varepsilon^{I}(\mbf{k},\lambda)a^{A}(\mbf{k},\lambda)$
are transverse, i.e. $\mbf{k}\cdot a^{A}(\mbf{k})=0$ and the polarization sum
is $D_{IJ}(\mbf{k})=\sum_{\lambda=1,2}\varepsilon_{I}(\mbf{k},\lambda)
\varepsilon_{J}(\mbf{k},\lambda)=\delta_{IJ}-\hat{k}_{I}\hat{k}_J$. 
In the massless basis, $m=0$, 
$u(\mbf{k})=v(\mbf{k})=(L(\mbf{k}),R(\mbf{k}))$ with the left Weyl
spinor $L(\mbf{k})=\left(-\sin(\Theta(\mbf{k})/2)\exp(-i\phi(\mbf{k})),
\cos(\Theta(\mbf{k})/2)\right)$.

Then, in the mean field approximation, the Hamiltonian Eq. (\ref{eq:2.1})
can be written in terms of creation/annihilation operators as  
\begin{eqnarray}
H &=& \int\frac{d\mbf{k}}{(2\pi)^3}~(k-\mu)~b^{\dagger}(\mbf{k})^{i}_{\alpha}
b(\mbf{k})^{i}_{\alpha}
+\int \frac{d\mbf{k}}{(2\pi)^3}~(\mu-k)~b^{\dagger}(\mbf{k})^{i}_{\alpha}
b(\mbf{k})^{i}_{\alpha}
\nonumber\\
&+&\int \frac{d\mbf{k}}{(2\pi)^3}~(k+\mu)~d^{\dagger}(\mbf{k})^{i}_{\alpha}
d(\mbf{k})^{i}_{\alpha}
+\int \frac{d\mbf{k}}{(2\pi)^3}~k~ a^{\dagger}(\mbf{k})a(\mbf{k})
\nonumber\\
&+&\int\frac{d\mbf{p}}{(2\pi)^3}~\Delta_{ij}^{\alpha\gamma}(\mbf{p})
T^{A}_{\alpha\beta}T^{A}_{\gamma\delta}
{\rm e}^{-i\phi(\mbf{p})}
\left(b(\mbf{p})^{i}_{\beta}b(-\mbf{p})^{j}_{\delta}
+d^{\dagger}(\mbf{p})^{i}_{\beta}d^{\dagger}(-\mbf{p})^{j}_{\delta}\right)
+c.c.
\nonumber\\
&+&\int\frac{d\mbf{k}}{(2\pi)^3}\frac{d\mbf{p}}{(2\pi)^3}
~\frac{3}{4}W(\mbf{k},\mbf{p}){\rm e}^{i\phi(\mbf{k})}
{\rm e}^{-i\phi(\mbf{p})}\nonumber\\
&\times&b^{\dagger}(\mbf{k})^{i}_{\alpha}
T^{A}_{\alpha\beta}b(\mbf{p})^{i}_{\beta}b^{\dagger}(-\mbf{k})^{j}_{\gamma}
T^{A}_{\gamma\delta}b(-\mbf{p})^{j}_{\delta}+\ldots
\nonumber\\
&+&\int\frac{d\mbf{k}}{(2\pi)^3}\frac{d\mbf{p}}{(2\pi)^3}
~g(\mbf{k},\mbf{p},\mbf{k}-\mbf{p})
\nonumber\\
&\times&b^{\dagger}(\mbf{k})^{i}_{\alpha}T^{A}_{\alpha\beta}
b(\mbf{p})^{i}_{\beta}\frac{a^{IA}(\mbf{k}-\mbf{p})}{2~|\mbf{k}-\mbf{p}|}
(u^{\dagger}(\mbf{k})\alpha^{I}u(\mbf{p}))+\ldots
\,,\label{eq:2.7}\end{eqnarray}
we explicitly display color $(\alpha,\beta)$, flavor $(i,j)$
and polarization $I$ indices. The first three terms describe
kinetic energies of particles, holes and antipaticles, respectively.
The diquark condensate $\Delta$ and the effective quark-quark interaction 
$W$ are unknown parameters, and dots show that there are other possible terms
which we ignored.  

Color-superconducting condensate $\Delta^{ij}_{\alpha\gamma}$
is a $N_c\times N_c$ matrix in fundamental color space 
$(\alpha,\gamma=1,...,N_c)$, and a $N_f\times N_f$ matrix in flavor 
space $(i,j=1,...,N_f)$. 
It can be parametrized for $N_f=3$ as \cite{AlfordRajagopalWilczek} 
\begin{eqnarray}
\Delta^{ij}_{\alpha\gamma}(\mbf{p})=3\left(\frac{1}{3}\left[\Delta_{8}(\mbf{p})
+\frac{1}{8}\Delta_{1}(\mbf{p})\right]\delta^{i}_{\alpha}\delta^{j}_{\gamma}
+\frac{1}{8}\Delta_{1}(\mbf{p})\delta^{i}_{\gamma}\delta^{j}_{\alpha}
\right)
\,,\label{eq:2.8}\end{eqnarray}
where $\Delta_1$ and $\Delta_8$ are the gaps in the singlet and octet
channels, respectively. We will also use the parametrization in terms
of symmetric, $\Delta_{(6,6)}$, and anti-symmetric,
$\Delta_{(\bar{3},\bar{3})}$, in color and flavor gap funtions,
$\Delta^{ij}_{\alpha\gamma}=\Delta_{(6,6)}(\delta^{i}_{\alpha}\delta^{j}_{\gamma}
+\delta^{i}_{\gamma}\delta^{j}_{\alpha})
+\Delta_{(\bar{3},\bar{3})}(\delta^{i}_{\alpha}\delta^{j}_{\gamma}
-\delta^{i}_{\gamma}\delta^{j}_{\alpha})$.The connection with the eigenvalue
gaps is given by $\Delta_{(\bar{3},\bar{3})}=1/2(\Delta_{8}-1/4\Delta_{1})$,
and $\Delta_{(6,6)}=1/2(\Delta_{8}+1/2\Delta_{1})$.

Replacing color $\alpha$ and flavor $i$
indices with a single color-flavor index $\rho$, 
$\Delta^{ij}_{\alpha\gamma}$ is diagonal in the CFL basis \cite{CFLbasis}
\begin{eqnarray}
b(\mbf{k})^{i}_{\alpha}=\sum_{\rho}\frac{\lambda^{\rho}_{i\alpha}}{\sqrt{2}}
~b(\mbf{k})^{\rho} 
\,,\label{eq:2.9}\end{eqnarray}
where $\lambda^{\rho}$ are the Gell-Mann matrices
for $\rho=1,...,8$ and $\lambda^{9}_{i\alpha}=\sqrt{2/3}\delta^{i}_{\alpha}$
for $\rho=9$, and strictly speaking we should have used different letter notation
for the CFL operators $b^{\rho}$. Then, 
$\Delta^{ij}_{\alpha\gamma}(\mbf{p})b(\mbf{p})^{i}_{\beta}
b(-\mbf{p})^{j}_{\delta}
T^{A}_{\alpha\beta}T^{A}_{\gamma\delta}=1/2\sum_{\rho}\Delta_{\rho}(\mbf{p})
b(\mbf{p})^{\rho}b(-\mbf{p})^{\rho}$, and the Hamiltonian Eq. (\ref{eq:2.7})
is given in the CFL basis by
\begin{eqnarray}
H &=& \sum_{\mbf{k},\rho}
|k-\mu|~b^{\dagger}_{\rho}(\mbf{k})b_{\rho}(\mbf{k})
+\sum_{\mbf{k},\rho}(k+\mu)~
d^{\dagger}_{\rho}(\mbf{k})d_{\rho}(\mbf{k})
+\sum_{\mbf{k}}~k~ a^{\dagger}(\mbf{k})a(\mbf{k})
\nonumber\\
&+&\frac{1}{2}\sum_{\mbf{p},\rho}\Delta_{\rho}(\mbf{p})
{\rm e}^{-i\phi(\mbf{p})}
\left(b_{\rho}(\mbf{p})b_{\rho}(-\mbf{p})
+d^{\dagger}_{\rho}(\mbf{p})d^{\dagger}_{\rho}(-\mbf{p})\right)+c.c.
\nonumber\\
&+&\frac{3}{4}\sum_{\mbf{k},\mbf{p},\rho,\rho'}
W^{\rho\rho'}(k,p)
\frac{1}{2}\lambda^{\rho}_{i\alpha}\lambda^{\rho}_{j\gamma}
\frac{1}{2}\lambda^{\rho'}_{i\beta}\lambda^{\rho'}_{j\delta}
T^{A}_{\alpha\beta}T^{A}_{\gamma\delta}
{\rm e}^{i\phi(\mbf{k})}{\rm e}^{-i\phi(\mbf{p})}
\nonumber\\
&\times&b^{\dagger}_{\rho}(\mbf{k})b_{\rho'}(\mbf{p})
b^{\dagger}_{\rho}(-\mbf{k})b_{\rho'}(-\mbf{p})
+\ldots
\nonumber\\
&+&\sum_{\mbf{k},\mbf{p},\rho,\rho'}
g^{\rho\rho'}(\mbf{k},\mbf{p},\mbf{k}-\mbf{p})
\frac{1}{2}\lambda^{\rho}_{i\alpha}T^{A}_{\alpha\beta}
\lambda^{\rho'}_{i\beta}
\nonumber\\
&\times&b^{\dagger}_{\rho}(\mbf{k})b_{\rho'}(\mbf{p})
\frac{a^{IA}(\mbf{k}-\mbf{p})}{2~|\mbf{k}-\mbf{p}|}
(u^{\dagger}(\mbf{k})\alpha^{I}u(\mbf{p}))
+\ldots
\,.\label{eq:2.10}\end{eqnarray}
where we used notation $\sum_{\mbf{k}}=\int d\mbf{k}/(2\pi)^3$.
As in \cite{AlfordRajagopalWilczek}, 
we change basis to creation/annihilation
operators $y$ and $z$ for quasiparticles (quasiholes) and 
quasiantiparticles, respectively,
\begin{eqnarray}
y_{\rho}(\mbf{k}) &=& \cos(\Theta_{\rho}^{y}(\mbf{k}))b^{\rho}(\mbf{k})
+\sin(\Theta_{\rho}^{y}(\mbf{k}))\exp(i\xi_{\rho}^{y}(\mbf{k}))
b^{\dagger}_{\rho}(-\mbf{k})
\nonumber\\
z_{\rho}(\mbf{k}) &=& \cos(\Theta_{\rho}^{z}(\mbf{k}))b^{\rho}(\mbf{k})
+\sin(\Theta_{\rho}^{z}(\mbf{k}))\exp(i\xi_{\rho}^{z}(\mbf{k}))
b^{\dagger}_{\rho}(-\mbf{k})
\,,\label{eq:2.11}\end{eqnarray}
with $y_{\rho}(\mbf{k})|0\rangle=z_{\rho}(\mbf{k})|0\rangle=0$.
In order to absorb the condensate term into a new free Hamiltonian,
we choose \cite{AlfordRajagopalWilczek}
$\cos(2\Theta_{\rho}^{y}(\mbf{k}))=|k-\mu|/E_{\rho}^{(-)}(\mbf{k})$,
$\cos(2\Theta_{\rho}^{z}(\mbf{k}))=|k+\mu|/E_{\rho}^{(+)}(\mbf{k})$,
and $\sin(2\Theta_{\rho}^{y}(\mbf{k}))
=\Delta_{\rho}(\mbf{k})/E_{\rho}^{(-)}(\mbf{k})$,
$\sin(2\Theta_{\rho}^{z}(\mbf{k}))
=\Delta_{\rho}(\mbf{k})/E_{\rho}^{(+)}(\mbf{k})$;
$\xi_{\rho}^{y}(\mbf{k})=\phi(\mbf{k})+\pi$,
$\xi_{\rho}^{z}(\mbf{k})=-\phi(\mbf{k})$.
As the result of this transformation, a new free Hamiltonian is given by
\begin{eqnarray}
\tilde{H}_{0} &=& \sum_{\mbf{k},\rho}E_{\rho}^{(-)}(\mbf{k})
~b^{\dagger}_{\rho}(\mbf{k})b_{\rho}(\mbf{k})
+\sum_{\mbf{k},\rho}E_{\rho}^{(+)}(\mbf{k})
~d^{\dagger}_{\rho}(\mbf{k})d_{\rho}(\mbf{k})
\nonumber\\
&+& \sum_{\mbf{k}}\omega(\mbf{k})~a^{\dagger}(\mbf{k})a(\mbf{k})
\,,\label{eq:2.12}\end{eqnarray}
where $E_{\rho}^{(-)}=\sqrt{(k-\mu)^2+\Delta_{\rho}(\mbf{k})^2}$,
$E_{\rho}^{(+)}=\sqrt{(k+\mu)^2+\Delta_{\rho}(\mbf{k})^2}$,
and $\omega(\mbf{k})=k$. The effective Hamiltonian Eq. (\ref{eq:2.10})
is given by
\begin{eqnarray} 
H = \tilde{H}_0+V_{qq}+V_{qg}
\,,\label{eq:2.13}\end{eqnarray}
with transformations of Eq. (\ref{eq:2.11}) made in the diquark 
interaction $V_{qq}$ and the quark-gluon coupling $V_{qg}$,
which are the final two terms in Eq. (\ref{eq:2.10}).
In the next section, our aim is to determine the unknown
parameters of the effective Hamiltonian Eq. (\ref{eq:2.13})
using flow equations.

\section{Flow equations}
\label{flowequations}

Flow equations are written for the unknown functions
$W^{\rho\rho'},g^{\rho\rho'},\Delta_{\rho}$ of the effective Hamiltonian
Eq. (\ref{eq:2.13}).
Calculations are performed using $y,z$ variables and the vacuum $|0\rangle$.
Similar calculations have been done in \cite{Gubankova}.
The first order flow equations $dV_{qg}/dl=[\eta,\tilde{H}_0]$,
$\eta=[\tilde{H}_0,V_{qg}]$ eliminate the quark-gluon coupling $V_{qg}$.
The generator of the transformation is given by
\begin{eqnarray}
\eta &=& \sum_{\mbf{k},\mbf{p},\rho,\rho'}
\eta^{\rho\rho'}(\mbf{k},\mbf{p},\mbf{k}-\mbf{p})
\frac{1}{2}\lambda^{\rho}_{i\alpha}T^{A}_{\alpha\beta}
\lambda^{\rho'}_{i\beta}
\nonumber\\
&\times&b^{\dagger}_{\rho}(\mbf{k})b_{\rho'}(\mbf{p})
\frac{a^{IA}(\mbf{k}-\mbf{p})}
{2~|\mbf{k}-\mbf{p}|}(u^{\dagger}(\mbf{k})\alpha^{I}u(\mbf{p}))
\,,\label{eq:3.1}\end{eqnarray}
then for the generator $\eta^{\rho\rho'}$ and coupling $g^{\rho\rho'}$
functions flow equations are written as
\begin{eqnarray}
\frac{dg^{\rho\rho'}}{dl} &=& -(E_{\rho}(\mbf{k})
-E_{\rho'}(\mbf{p})-\omega(\mbf{k}-\mbf{p}))^2~g^{\rho\rho'}
\nonumber\\
\eta^{\rho\rho'} &=& (E_{\rho}(\mbf{k})
-E_{\rho'}(\mbf{p})-\omega(\mbf{k}-\mbf{p}))~g^{\rho\rho'}
\,.\label{eq:3.2}\end{eqnarray}
The solution of Eq. (\ref{eq:3.2}) reads
\begin{eqnarray}
g^{\rho\rho'}(l) &=&g(0)\exp\left(-(E_{\rho}(\mbf{k})
-E_{\rho'}(\mbf{p})-\omega(\mbf{k}-\mbf{p}))^2~l\right)
\,,\label{eq:3.3}\end{eqnarray}
where $g(0)=g$ is the bare coupling constant.
As $l\rightarrow\infty$ the coupling is eliminated as long
as the states in the exponent are not degenerate.

Using the generator Eq. (\ref{eq:3.1}) and the quark-gluon coupling
Eq. (\ref{eq:2.10}), we write the second order flow equations  
for the effective quark-quark interaction,
$dV_{qq}/dl=[\eta,V_{qg}]_{two-body}$, and for the diquark self-energy,
$d\Sigma_{q}/dl=[\eta,V_{qg}]_{one-body}$. 
Flow equation for the diquark interaction is given by 
\begin{eqnarray} 
\frac{dV_{qq}}{dl} &=& -\sum_{\mbf{k},\mbf{p},\rho\,rho'}
b^{\dagger}_{\rho}(\mbf{k})b^{\dagger}_{\rho}(-\mbf{k})
b_{\rho'}(\mbf{p})b_{\rho'}(-\mbf{p})
\frac{1}{2}\lambda^{\rho}_{i\alpha}\lambda^{\rho}_{j\gamma} 
\frac{1}{2}\lambda^{\rho'}_{i\beta}\lambda^{\rho'}_{j\delta} 
T^{A}_{\alpha\beta}T^{A}_{\gamma\delta}  
\nonumber\\
&&\hspace{-1.2cm}(u^{\dagger}(\mbf{k})\alpha^{I}u(\mbf{p}))
(u^{\dagger}(-\mbf{k})\alpha^{J}u(\mbf{-p}))
\frac{D_{IJ}(\mbf{k}-\mbf{p})}{2\omega(\mbf{k}-\mbf{p})}
\label{eq:3.4}\\
&&\hspace{-1.2cm}
\left(\eta^{\rho\rho'}(\mbf{k},\mbf{p},\mbf{k}-\mbf{p})
g^{\rho'\rho}(\mbf{p},\mbf{k},\mbf{k}-\mbf{p})
+\eta^{\rho'\rho}(\mbf{p},\mbf{k},\mbf{k}-\mbf{p})
g^{\rho\rho'}(\mbf{k},\mbf{p},\mbf{k}-\mbf{p})\right)\nonumber
\,.\end{eqnarray}
Integrating the flow equation Eq. (\ref{eq:3.4})
over $l=[0,\infty)$, with the final value $V_{qq}=V_{qq}(l=\infty)$, 
and adding the instantaneous interaction Eq. (\ref{eq:2.3}), 
we get 
\begin{eqnarray} 
V_{qq} &=&- \sum_{\mbf{k},\mbf{p},\rho,\rho'}
b^{\dagger}_{\rho}(\mbf{k})b^{\dagger}_{\rho}(-\mbf{k})
b_{\rho'}(\mbf{p})b_{\rho'}(-\mbf{p}) 
\frac{1}{2}\lambda^{\rho}_{i\alpha}\lambda^{\rho}_{j\gamma} 
\frac{1}{2}\lambda^{\rho'}_{i\beta}\lambda^{\rho'}_{j\delta} 
T^{A}_{\alpha\beta}T^{A}_{\gamma\delta}
\nonumber\\
&&\hspace{-1cm}\left(V^{\rho\rho'}(\mbf{k},\mbf{p}) 
(u^{\dagger}(\mbf{k})\alpha^{I}u(\mbf{p}))
(u^{\dagger}(-\mbf{k})\alpha^{J}u(\mbf{-p}))D_{IJ}(\mbf{k}-\mbf{p})\right.
\nonumber\\
&&\hspace{-1cm}\left.
+V^{\phantom {\rho\rho'}}\hspace{-0.4cm}(\mbf{k},\mbf{p})
(u^{\dagger}(\mbf{k})u(\mbf{p}))(u^{\dagger}(-\mbf{k})u(\mbf{-p}))
\right)
\,,\label{eq:3.5}\end{eqnarray}
with
\begin{eqnarray} 
V^{\rho\rho'}(\mbf{k},\mbf{p}) &=& -\frac{g^2}{2}
\frac{1}{(E^{\rho}(\mbf{k})-E^{\rho'}(\mbf{p}))^2+\omega_{M}(\mbf{k}-\mbf{p})^2}
\nonumber\\
V(\mbf{k},\mbf{p}) &=& \frac{g^2}{2}
\frac{1}{\omega_{E}(\mbf{k}-\mbf{p})^2}
\,,\label{eq:3.6}\end{eqnarray}
where we have used solutions for $\eta^{\rho\rho'}$
and $g^{\rho\rho'}$, Eqs. (\ref{eq:3.2},\ref{eq:3.3}).
Energies for magnetic and electric gluons, $\omega_{M}$
and $\omega_{E}$ respectively, are specified further. 
Keeping only left-left components, matrix elements are given by
\begin{eqnarray}
&&\hspace{-1.2cm}(u^{\dagger}(\mbf{k})\alpha^{I}u(\mbf{p}))
(u^{\dagger}(-\mbf{k})\alpha^{J}u(\mbf{-p}))D_{IJ}(\mbf{k}-\mbf{p})=
\nonumber\\
&&(-1){\rm e}^{-i\phi(\mbf{p})}{\rm e}^{i\phi(\mbf{k})}
\left(-\frac{3-\hat{k}\cdot\hat{p}}{2}+\frac{1-\hat{k}\cdot\hat{p}}{2}
\frac{(k+p)^2}{(\mbf{k}-\mbf{p})^2}\right)
\nonumber\\
&&\hspace{-1.2cm}(u^{\dagger}(\mbf{k})u(\mbf{p}))
(u^{\dagger}(-\mbf{k})u(\mbf{-p})) =
(-1){\rm e}^{-i\phi(\mbf{p})}{\rm e}^{i\phi(\mbf{k})}
\left(\frac{1+\hat{k}\cdot\hat{p}}{2}\right)
\,,\label{eq:3.7}\end{eqnarray}
where we used Fierz transform 
$(\bar{u}(\mbf{k})O^{A}u(\mbf{p}))(\bar{u}(-\mbf{k})O^{B}u(-\mbf{p}))
=\epsilon\sum_{CD}f^{AB}_{CD}(\bar{u}_{C}(\mbf{p})O^{C}u(-\mbf{p}))
(\bar{u}(-\mbf{k})O^{D}u_{C}(\mbf{k}))$, where $\epsilon=-1$
for $O^{A},O^{B}=\gamma^{\mu}$, and $C\bar{u}^{T}=u_C$,
$u^{T}C=\bar{u}_{C}$. For the above matrix elements
the coefficients $f$ in Fierz transform $O^{A}\times O^{B}\rightarrow
O^{C}\times O^{D}$ are given by 
$\gamma_0\times \gamma_0\rightarrow 1/4(1\times 1-\gamma_5\times\gamma_5)
+1/4\sigma_{\mu\nu}\times \sigma^{\mu\nu}$, and
$\gamma_i\times\gamma^{j}\rightarrow \delta_{i}^{j}1/4(1\times 1-
\gamma_5\times\gamma_5)+\delta_{i}^{j}1/4\sigma_{\mu\nu}\times\sigma^{\mu\nu}
+1/4(\sigma_{i\mu}\times\sigma^{j\mu}+\sigma_{\mu i}\times\sigma^{\mu j}
-\sigma_{i\mu}\times\sigma^{\mu j}-\sigma_{\mu i}\times\sigma^{j\mu})$,
where $\mu,\nu=(0,i=1,2,3)$, $\sigma_{\mu\nu}=1/2[\gamma_{\mu},\gamma_{\nu}]$.
The diquark condensates are $1\times1$-pseudoscalar $(\bar{u}_{C}u)$,
$\gamma_5\times\gamma_5$-scalar $(\bar{u}_{C}\gamma u)$, and
$\sigma_{\mu\nu}\times\sigma^{\mu\nu}$-vector 
$(\bar{u}_{C}\sigma_{\mu\nu}u)$.
Performing algebra with $\lambda$-matrices in Eq. (\ref{eq:3.5}), 
we get
\begin{eqnarray}
V_{qq}&=&- \sum_{\mbf{k},\mbf{p}}
(-1){\rm e}^{-i\phi(\mbf{p})}{\rm e}^{i\phi(\mbf{k})}
\nonumber\\
&&\hspace{-1cm} \left(b^{\dagger}_{8}(\mbf{k})b^{\dagger}_{8}(-\mbf{k})
b_{1}(\mbf{p})b_{1}(-\mbf{p})F^{81}(\mbf{k},\mbf{p})
+b^{\dagger}_{1}(\mbf{k})b^{\dagger}_{1}(-\mbf{k})
b_{8}(\mbf{p})b_{8}(-\mbf{p})F^{18}(\mbf{k},\mbf{p})\right.
\nonumber\\
&-&\left.b^{\dagger}_{8}(\mbf{k})b^{\dagger}_{8}(-\mbf{k})
b_{8}(\mbf{p})b_{8}(-\mbf{p})2F^{88}(\mbf{k},\mbf{p})\right)  
\,,\label{eq:3.8}\end{eqnarray}
where we introduced
\begin{eqnarray} 
F^{\rho\rho'}=V_{dyn}^{\rho\rho'}(\mbf{k},\mbf{p})
\left(\frac{3-\hat{k}\cdot\hat{p}}{2}\right)
+V_{inst}(\mbf{k},\mbf{p})
\left(\frac{1+\hat{k}\cdot\hat{p}}{2}\right)
\,,\label{eq:3.9}\end{eqnarray}
and
\begin{eqnarray} 
V_{dyn}^{\rho\rho'}(\mbf{k},\mbf{p}) &=& \frac{2g^2}{3}
\frac{1}{(E^{\rho}(\mbf{k})-E^{\rho'}(\mbf{p}))^2
+\omega_{M}(\mbf{k}-\mbf{p})^2}
\nonumber\\
V_{inst}(\mbf{k},\mbf{p}) &=& \frac{2g^2}{3}
\frac{1}{\omega_{E}(\mbf{k}-\mbf{p})^2}
\,.\label{eq:3.10}\end{eqnarray}
Dynamically generated by flow equations interaction describes
magnetic gluon exchange, and the instantaneous interaction
-electric gluon exchange.
Comparing Eq. (\ref{eq:3.8}) with Eq. (\ref{eq:2.10}) for $V_{qq}$, 
the effective quark-quark interaction is given by
\begin{eqnarray}  
W^{\rho\rho'}(\mbf{k},\mbf{p})&=& -F^{\rho\rho'}(\mbf{k},\mbf{p})
=-V_{dyn}^{\rho\rho'}(\mbf{k},\mbf{p})-V_{inst}^{\rho\rho'}(\mbf{k},\mbf{p})
\,,\label{eq:3.11}\end{eqnarray}
in the collinear limit $\hat{k}\cdot\hat{p}=1$.
This interaction is attractive in the singlet and octet channels.
Note that $W^{11}=0$, and condensation in singlet channel
is driven by $W^{81}$.

Dynamical magnetic interaction, Eq. (\ref{eq:3.10}), has 
the form $-1/(\mbf{q}^2+\delta E^2)$ instead of $-1/(\mbf{q}^2-\delta E^2)$,
that is produced by the equal time perturbation theory, where
$\delta E$ is the enegry difference of in- and out-going quarks. 
The latter has the pole,therefore usually
$\delta E$ is neglected near the Fermi surface.
Our interaction is regular, and $\delta E$ will play an important 
role.
 
In order to incorporate effects of the dense quark medium, 
we include polarization operators for electric and magnetic gluons, 
modifying the gluon single energy as done in the work of Pisarski and Rischke
\cite{N_f=2} 
\begin{eqnarray} 
\frac{1}{\omega_{M}(\mbf{k}-\mbf{p})^2} &=&\frac{1}{2}\left(
\frac{(\mbf{k}-\mbf{p})^4}{(\mbf{k}-\mbf{p})^6+M^4(E(\mbf{k})+E(\mbf{p}))^2)}
\right.
\nonumber\\
&+&\left.
\frac{(\mbf{k}-\mbf{p})^4}{(\mbf{k}-\mbf{p})^6+M^4(E(\mbf{k})-E(\mbf{p}))^2)}
\right)
\nonumber\\
\frac{1}{\omega_{E}(\mbf{k}-\mbf{p})^2} &=&
\frac{1}{(\mbf{k}-\mbf{p})^2+3m_g^2}
\,,\label{eq:3.12}\end{eqnarray}
where $M^2=(3\pi/4)m_g^2$, $m_g^2=N_fg^2\mu^2/(6\pi)^2$.
Eq. (\ref{eq:3.12}) is used in the complete diquark interaction
Eq. (\ref{eq:3.10}).

Quark self-energy has $b^{\dagger}b$ and $bb+b^{\dagger}b^{\dagger}$
components associated with normal and anomalous propagation,
respectively. 
For the anomalous propagation self-energy the flow equation 
is given by
\begin{eqnarray} 
\frac{d\Sigma_{qq}}{dl} &=& -\sum_{\mbf{k},\mbf{p},\rho,\rho'}
\frac{1}{2}\lambda^{\rho}_{i\alpha}\lambda^{\rho}_{j\gamma} 
\frac{1}{2}\lambda^{\rho'}_{i\beta}\lambda^{\rho'}_{j\delta} 
T^{A}_{\alpha\beta}T^{A}_{\gamma\delta}
\nonumber\\
&&\hspace{-1.2cm}
\left(<b^{\dagger}_{\rho}(\mbf{k})b^{\dagger}_{\rho}(-\mbf{k})>
b_{\rho'}(\mbf{p})b_{\rho'}(-\mbf{p})
+<b_{\rho'}(\mbf{p})b_{\rho'}(-\mbf{p})>
b^{\dagger}_{\rho}(\mbf{k})b^{\dagger}_{\rho}(-\mbf{k})\right)
\nonumber\\
&& (u^{\dagger}(\mbf{k})\alpha^{I}u(\mbf{p}))
(u^{\dagger}(-\mbf{k})\alpha^{J}u(\mbf{-p}))
\frac{D_{IJ}(\mbf{k}-\mbf{p})}{2\omega(\mbf{k}-\mbf{p})}
\nonumber\\
&&\hspace{-1.2cm}\left(\eta^{\rho\rho'}(\mbf{k},\mbf{p},\mbf{k}-\mbf{p})
g^{\rho'\rho}(\mbf{p},\mbf{k},\mbf{k}-\mbf{p})
+\eta^{\rho'\rho}(\mbf{p},\mbf{k},\mbf{k}-\mbf{p})
g^{\rho\rho'}(\mbf{k},\mbf{p},\mbf{k}-\mbf{p})\right)
\nonumber\\
&+&(Terms\sim <dd>,<d^{\dagger}d^{\dagger}>)
\,.\label{eq:3.13}\end{eqnarray}
We neglect the contribution of antiparticles, since
it is regular near the Fermi surface,
$<dd>\sim \sin(2\Theta^{z})\sim \Delta/2\mu$,
and does not change much the condensate of particles.
The anomalous propagator for particles is given by
\begin{eqnarray}
<b^{\dagger}_{\rho}(\mbf{k})b^{\dagger}_{\rho'}(-\mbf{k})>
&=&-\frac{1}{2}\sin(2\Theta_{\rho}(\mbf{k}))\exp(-i\xi_{\rho}(\mbf{k}))
\,,\label{eq:3.14}\end{eqnarray}
where the average is calculated in $|0\rangle$ vacuum.

Integrating over $l=[0,1/\lambda^2]$, where $\lambda$
is the UV cut-off, and adding the self-energy instantaneous term 
comming from normal-ordering Eq. (\ref{eq:2.3}) in the $|0\rangle$ vacuum, 
we get 
\begin{eqnarray} 
\Sigma_{qq}(\lambda) &=& \sum_{\mbf{k},\mbf{p},\rho,\rho'}
{\frac{1}{2}\lambda^{\rho}_{i\alpha}\lambda^{\rho}_{j\gamma} 
\frac{1}{2}\lambda^{\rho'}_{i\beta}\lambda^{\rho'}_{j\delta} 
T^{A}_{\alpha\beta}T^{A}_{\gamma\delta}} 
<b^{\dagger}_{\rho}(\mbf{k})b^{\dagger}_{\rho}(-\mbf{k})>
b_{\rho'}(\mbf{p})b_{\rho'}(-\mbf{p}) 
\nonumber\\
&&\hspace{-1cm}\left(V^{\rho\rho'}(\mbf{k},\mbf{p}) 
(u^{\dagger}(\mbf{k})\alpha^{I}u(\mbf{p}))
(u^{\dagger}(-\mbf{k})\alpha^{J}u(\mbf{-p}))D_{IJ}(\mbf{k}-\mbf{p})\right.
\nonumber\\
&&\hspace{-1cm}\left.+V^{\phantom {\rho\rho'}}\hspace{-0.4cm}(\mbf{k},\mbf{p})
(u^{\dagger}(\mbf{k})u(\mbf{p}))(u^{\dagger}(-\mbf{k})u(\mbf{-p}))
\right)R^{\rho\rho'}(\mbf{k},\mbf{p};\lambda) +c.c.
\,,\label{eq:3.15}\end{eqnarray}
where $V^{\rho,\rho'}$ and $V$ are given in Eq. (\ref{eq:3.6}), and
the the UV regulator is given by
\begin{eqnarray} 
R^{\rho\rho'}(\mbf{k},\mbf{p};\lambda) &=& 
\exp\left(-[(E^{\rho}(\mbf{k})-E^{\rho'}(\mbf{p}))^2
+\omega_{M}(\mbf{k}-\mbf{p})^2]/\lambda^2\right)
\,,\label{eq:3.16}\end{eqnarray}
Change in sign appear due to adopting conventional scaling up as in 
the perturbation theory, instead of scaling down in flow equations.
Performing calculations with $\lambda$-matrices, we have
\begin{eqnarray}
\Sigma_{qq}&=&\sum_{\mbf{k},\mbf{p}}
(-1){\rm e}^{-i\phi(\mbf{p})}{\rm e}^{i\phi(\mbf{k})}
\left(b_{1}(\mbf{p})b_{1}(-\mbf{p})
<b^{\dagger}_{8}(\mbf{k})b^{\dagger}_{8}(-\mbf{k})>
F^{81}(\mbf{k},\mbf{p})\right.
\nonumber\\
&&\hspace{-1.2cm}\left.+b_{8}(\mbf{p})b_{8}(-\mbf{p})
[<b^{\dagger}_{1}(\mbf{k})b^{\dagger}_{1}(-\mbf{k})>
F^{18}(\mbf{k},\mbf{p})
-<b^{\dagger}_{8}(\mbf{k})b^{\dagger}_{8}(-\mbf{k})>
2F^{88}(\mbf{k},\mbf{p})]\right)
\nonumber\\
&\times&R^{\rho\rho'}(\mbf{k},\mbf{p};\lambda)
+c.c. 
\,.\label{eq:3.17}\end{eqnarray}
Comparing Eq. (\ref{eq:3.17}) with Eq. (\ref{eq:2.10})
for the gap functions, we obtain the system of gap equations 
\begin{eqnarray}
\Delta_1(\mbf{p}) &=&8G^{81}(\mbf{p})
\nonumber\\
\Delta_8(\mbf{p}) &=&G^{18}(\mbf{p})-2G^{88}(\mbf{p})
\,,\label{eq:3.18}\end{eqnarray}
where
\begin{eqnarray}
G^{\rho\rho'}(\mbf{p})&=&-\frac{1}{4}\int \frac{d\mbf{k}}{(2\pi)^3}
\frac{1}{2}\sin(2\Theta_{\rho}(\mbf{k}))
F^{\rho\rho'}(\mbf{k},\mbf{p})
R^{\rho\rho'}(\mbf{k},\mbf{p};\lambda)
\,,\label{eq:3.19}\end{eqnarray}
with $F^{\rho\rho'}$ is given by Eq. (\ref{eq:3.9}),
and $R^{\rho\rho'}$ by Eq. (\ref{eq:3.16}).

\section{Solving the system of gap equations}
\label{gapequation}

We solve approximately the system of gap equations Eq. (\ref{eq:3.18}) 
in two limiting cases of the Pipperd and London type superconductor.
Magnetic interaction Eq. (\ref{eq:3.10}) is regulated by
the term generated by flow equations, with dispersion $q\sim E$, 
and by the magnetic gluon polarization 
operator, Eq. (\ref{eq:3.12}), with dispersion $q\sim E^{1/3}$.
When the first term dominates, superconductor is of London type,
the second term-Pipperd type.

We take instead of the smooth regulator $R$, Eq. (\ref{eq:3.16}), 
a sharp cut-off, $|k-\mu|\leq\delta$.
Near the Fermi surface the integration measure is given by
$\int d\mbf{k}=2\pi\mu^2\int_{-\delta}^{\delta}
d(k-\mu)\int_{-1}^{1}d\cos\theta$, and since the integral is even
in $(k-\mu)$ one has $\int_{\delta}^{\delta}\rightarrow 2\int_{0}^{\delta}$.
In the denominator of $F$, Eq. (\ref{eq:3.9}), we approximate 
$(\mbf{k}-\mbf{p})^2=2\mu^2(1-\cos\theta)$.

Performing the $\theta$-integration, we get for the
Pipperd superconductor:
\begin{eqnarray}
G^{\rho\rho'}(\mbf{p}) &=& -\frac{1}{4}\bar{g}^2
\int_{0}^{\delta}
\frac{d(k-\mu)}{E_{\rho}(\mbf{k})}\Delta_{\rho}(\mbf{k})
\frac{1}{2}\ln\left(\frac{b^2\mu^2}{|E(\mbf{k})^2-E(\mbf{p})^2|}
\right)
\,,\label{eq:4.1}\end{eqnarray}
where
\begin{eqnarray}
\bar{g} &=&\frac{g}{3\sqrt{2}\pi}~;~
b_{M} = 32\pi\left(\frac{2}{N_fg^2}\right)~,~
b_{M+E} = 256\pi^4\left(\frac{2}{N_fg^2}\right)^{5/2}
\,,\label{eq:4.2}\end{eqnarray}
and superscript $M$ and $M+E$ means that only magnetic
or magnetic and electric components are taken.
For the London superconductor:
\begin{eqnarray}
G^{\rho\rho'}(\mbf{p}) &=& -\frac{1}{4}\bar{g}^2
\int_{0}^{\delta}
\frac{d(k-\mu)}{E_{\rho}(\mbf{k})}\Delta_{\rho}(\mbf{k})
\frac{1}{2}\ln\left(\frac{b^2\mu^2}
{(E_{\rho}(\mbf{k})-E_{\rho'}(\mbf{p}))^2}\right)
\,,\label{eq:4.3}\end{eqnarray}
where
\begin{eqnarray}
\bar{g} &=&\frac{g}{\sqrt{6}\pi}~;~
b_{M} = 2~,~
b_{M+E} = 4\pi\left(\frac{2}{N_fg^2}\right)^{1/2}
\,.\label{eq:4.4}\end{eqnarray}
Substituting $G^{\rho\rho'}$ Eqs. (\ref{eq:4.1},\ref{eq:4.3})
into Eq. (\ref{eq:3.18}), we have the system of Eliashberg
type of equations. Different effective coupling $\bar{g}$
is obtained because of different dispersion law
in the two cases.
The coupling $\bar{g}$ defines the exponent of the solution
for the gap $\Delta$ and $b$ the preexponential factor.

If we are interested only in the correct exponential dependence,
we can neglect $\rho$ dependence by the quark energies
in the anomalous quark propagator ($\sin$ factor in G) and in the 
gluon propagator (under the $\ln$). Indeed, when estimating
the momentum-independent gap, double logarithm arises from the region
$\Delta\ll k\ll \mu$ where the integral reduces to
\begin{eqnarray} 
\Delta\sim\bar{g}^2\int_{\Delta}^{\delta}\frac{d(k-\mu)}{(k-\mu)}
\ln\left(\frac{b\mu}{|k-\mu|}\right)\Delta
\,,\label{eq:4.5}\end{eqnarray}
with the solution $\Delta\sim b\mu\exp(-\sqrt{2}/\bar{g})$;
i.e. double logarithm responsible for the exponential solution
does not depend on the quark energies. To calculate
the preexponential factor correctly, one should make the rescaling
in the integral as suggested by Sch\"afer \cite{N_f=3}.

In order to convert the integral gap equation into differential one,
we split the logarithm in $G^{\rho\rho'}$ as suggested by Son
\cite{Son}
\begin{eqnarray}
\ln\left(\frac{b\mu}{E(\mbf{p})}\right)\theta(p-k)
+\ln\left(\frac{b\mu}{E(\mbf{k})}\right)\theta(k-p)
\,,\label{eq:4.6}\end{eqnarray}
that gives the same expression in two cases, Eq. (\ref{eq:4.1})
and Eq. (\ref{eq:4.3}).
Coupled gap equations Eq. (\ref{eq:3.18}) decouple
for the antitriplet and sixtet in color and flavor gaps,
introduced after Eq. (\ref{eq:2.8}), 
\begin{eqnarray}
\Delta_{(\bar{3},\bar{3})}(\mbf{p}) &=& \bar{g}^2
\ln\left(\frac{b\mu}{E(\mbf{p})}\right)
\int_{0}^{(p-\mu)}\frac{d(k-\mu)}{E(\mbf{k})}
\Delta_{(\bar{3},\bar{3})}(\mbf{k})
\nonumber\\
&+&\bar{g}^2
\int_{(p-\mu)}^{\delta}\frac{d(k-\mu)}{E(\mbf{k})}
\ln\left(\frac{b\mu}{E(\mbf{k})}\right)
\Delta_{(\bar{3},\bar{3})}(\mbf{k})
\nonumber\\
\Delta_{(6,6)}(\mbf{p}) &=&
-\frac{\bar{g}^2}{2} 
\ln\left(\frac{b\mu}{E(\mbf{p})}\right)
\int_{0}^{(p-\mu)}\frac{d(k-\mu)}{E(\mbf{k})}
\Delta_{(6,6)}(\mbf{k})
\nonumber\\
&-&\frac{\bar{g}^2}{2}
\int_{(p-\mu)}^{\delta}\frac{d(k-\mu)}{E(\mbf{k})}
\ln\left(\frac{b\mu}{E(\mbf{k})}\right)
\Delta_{(6,6)}(\mbf{k})
\,.\label{eq:4.7}\end{eqnarray}
Introducing the variable, as by Pisarski and Rischke \cite{N_f=2},
\begin{eqnarray}
x &=&\ln\left(\frac{2b\mu}{p-\mu+E(\mbf{p})}\right)
\,,\label{eq:4.8}\end{eqnarray}
the integratiom measure is given by $d(k-\mu)/E(\mbf{k})=dx/x$,
and the integral equations, Eq. (\ref{eq:4.7}),
reduce to differential equations
\begin{eqnarray}
\frac{d^2}{dx^2}\Delta_{(\bar{3},\bar{3})}
+\bar{g}^2\Delta_{(\bar{3},\bar{3})} &=& 0
\nonumber\\
\frac{d^2}{dx^2}\Delta_{(6,6)}
-\frac{\bar{g}^2}{2}\Delta_{(6,6)} &=& 0
\,,\label{eq:4.9}\end{eqnarray}
with initial conditions at the Fermi surface 
$d\Delta/dx(x=x_0)=0$ and $\Delta(x=x_0)=\Delta_0$;
also away from the Fermi surface $\Delta(x=0)=0$.
The negative sign in the second equation means the repulsion
in the sixtet channel. As function of momentum there is a trivial 
solution in the sixtet channel. In the antitriplet channel,
the solution reads
\begin{eqnarray}
\Delta_{(\bar{3},\bar{3})} &=& \Delta_{0}\sin(\bar{g}x)~,~
\Delta_{0} = 2b\mu\exp(-\frac{\pi}{2\bar{g}})
\,,\label{eq:4.10}\end{eqnarray}
where $\bar{g}$ and $b$ are given in Eqs. (\ref{eq:4.2},\ref{eq:4.4}).

\section{Conclusions}
\label{conclusion}

Using flow equations, we derived an effective quark-quark 
interaction and obtained the coupled set of gap equations
for the condensates of the CFL phase of massless $N_f=3$
dense QCD. Diquark interaction, generated dynamically by
flow equations, is a long-range magnetic gluon exchange
regulated by two different sources in the infrared region.
One term describes the retardation effects of a magnetic gluon
caused by Landau damping in dense quark-gluon plasma (normal phase).
At small temperature, far below the melting of superconducting 
condensate $T\ll T_0$, Landau damping is the dominant mechanism 
in the infrared, that corresponds to the Pipperd type of superconductor.
The other term describes retardation due to propagating of a magnetic
gluon in a superconducting matter, i.e. through the multiple 
scattering of a gluon at diquark Cooper pairs. This mechanism
is dominant at temperatures close to a melting point of 
superconducting condensate, $T-T_0\ll T_0$, and corresponds
to the London type of superconductor.

We obtain approximate analytical solutions of the gap equations
in these two limiting cases. The dominant contribution to the
condensate comes in the color antitriplet, flavor antitriplet
channel. The color sextet, flavor sextet contribution is small
but non-zero. In the color and flavor antisymmetric channel,
we obtain parametric enhacement of the London type condensate
\begin{eqnarray}
\Delta_{0}\sim\exp(-\sqrt{\frac{3}{2}}\frac{\pi^2}{g})
\,,\label{eq:5.1}\end{eqnarray}
versus the Pipperd type condensate 
\begin{eqnarray}
\Delta_{0}\sim\exp(-\frac{3}{\sqrt{2}}\frac{\pi^2}{g})
\,.\label{eq:5.2}\end{eqnarray}
The same conclusion holds for $N_f=2$.
Numerically, the $(\bar{3},\bar{3})$ codensates
are almost the same for the London and Pipperd type 2SC 
superconductors, $N_f=2$. However, the condensate of the London type 
superconductor is sufficiently larger than of the Pipperd superconductor
in the CFL phase, $N_f=3$. This has important implications,
which regime is energetically favorable and will actually be realized in
neutron stars.

Independence of the gap on number of flavors for magnetic field
and slow dependence on $N_f$ for the sum of magnetic and electric fields
in London type superconductors works in favor that the CFL phase
will be observed in neutron stars, and that there is no
window for the 2SC phase \cite{Alford}.

Our calculations have been done at zero temperature. Rigorous
calculations of the condensates which include calculations of
the penetration depth of magnetic field and coherence length 
at non-zero temperature are necessary. Qualitatively it will not 
affect our results, since our conclusions rely on similar trends 
known in metal superconductors, which are derived from Landau-Ginzburg
theory and are proven experimentally \cite{Abrikosov}.  

We included all possible condensates, $<qCq>$, $<qC\gamma_{5}q>$ and 
$<qC\sigma_{\mu\nu}q>$, which do not mix left and right components,
that accounted for the angle dependence in the gap equations.
It is interesting to add the instanton-induced 't Hooft interaction
and to consider simultaneously the chiral condensation in 
$\bar{q}q$ channel. Generalization to a non-zero strange quark mass,
and adding kaon condensates is important for neutron star phenomenology.


\begin{thebibliography}{0}

\bibitem{old} B.~C.~Barrois, 
{\it Nucl. Phys. } {\bf B129}, 390 (1977);
S.~C.~Frautschi,
in {\it Hadronic Matter at Extreme Energy Density}, edited by 
N.~Cabibbo and L.~Sertorio (Plenum, New York, 1980);
D.~Bailin and A.~Love,
{\it Phys. Rep.}  {\bf 107}, 325 (1984).

\bibitem{revised} M.~G.~Alford, K.~Rajagopal and F.~Wilczek,
{\it Phys. Lett.}  {\bf B422}, 247 (1998);
R.~Rapp, T.~Sch\"{a}fer, E.~V.~Shuryak and M.~Velkovsky,
{\it Phys. Rev. Lett.}  {\bf 81}, 53 (1998).

\bibitem{AlfordRajagopalWilczek} M.~Alford, K.~Rajagopal and F.~Wilczek,
{\it Nucl. Phys.}  {\bf B537}, 443 (1999);

\bibitem{Son} D.~T.~Son,
{\it Phys. Rev.}  {\bf D59}, 094019 (1999);
R.~D.~Pisarski and D.~H.~Rischke,
{\it Phys. Rev. Lett.}  {\bf 83}, 37 (1999).

\bibitem{N_f=2} T.~Schafer and F.~Wilczek,
{\it Phys. Rev.}  {\bf D60}, 114033 (1999);
D.~K.~Hong, V.~A.~Miransky, I.~A.~Shovkovy and L.~C.~R.~Wijewardhana,
{\em ibid.} {\bf D61}, 056001 (2000);
{\bf D62}, 059903(E) (2000);
R.~D.~Pisarski and D.~H.~Rischke,
{\em ibid.} {\bf D61}, 051501 (2000);
{\em ibid.} {\bf D61} 074017 (2000). 

\bibitem{N_f=2more} S.~D.~Hsu and M.~Schwetz,
{\it Nucl. Phys.}  {\bf B572}, 211 (2000);
W.~E.~Brown, J.~T.~Liu and H.~C.~Ren,
{\it Phys. Rev.}  {\bf D61}, 114012 (2000).

\bibitem{N_f=3} K.~Rajagopal and E.~ Shuster,
{\it Phys. Rev.}  {\bf D62}, 085007 (2000);
I.~A.~Shovkovy and L.~C.~R.~Wijewardhana,
{\it Phys. Lett.}  {\bf B470}, 189 (1999);
T.~Sch\"{a}fer,
{\it Nucl. Phys.} {\bf B575}, 269 (2000).

\bibitem{Gubankova} E.~Gubankova, hep-ph/0112213.

\bibitem{CFLbasis} D.~T.~Son and M.~A.~Stephanov,
{\it Phys. Rev.}  {\bf D61}, 074012 (2000);
{\bf D62}, 059902(E) (2000).

\bibitem{Wegner} F.~Wegner, {\it Ann. Phys. (Leipzig)} 
{\bf 3}, 77 (1994);
{\it Physics Reports} {\mbf 348}, 77 (2001).

\bibitem{LenzWegner} P.~Lenz and F.~Wegner,
cond-mat/9604087; for a recent work see
I.~Grote, E.~Koerding, F.~Wegner, cond-mat/0106604.

\bibitem{Abrikosov} A.~A.~Abrikosov, L.~P.~Gorkov, and
I.~E.~Dzyaloshinski, ``Methods of quantum field theory
in statistical physics,'' Dover publications, Inc., New York.
 
\bibitem{Alford}
M.~Alford and K.~Rajagopal, {\it JHEP} {\bf 0206}, 031 (2002). 








\end{thebibliography}
\end{document}